\documentclass[twocolumn,aps,prd,amsmath,amssymb,floatfix,amsmath,superscriptaddress]{revtex4-2}

\usepackage{commath}
\usepackage{units}
\usepackage{graphicx}% Include figure files
\usepackage{dcolumn}% Align table columns on decimal point
\usepackage{bm}% bold math
\usepackage{braket}
\usepackage{color,epsfig}
\usepackage{slashed}
\usepackage{hyperref}
\usepackage{subfigure}
\usepackage{feynmf}
\usepackage{calrsfs}

\usepackage{booktabs}

\newcommand{\bea}{\begin{eqnarray}}
\newcommand{\eea}{\end{eqnarray}}
\newcommand{\be}{\begin{equation}}
\newcommand{\ee}{\end{equation}}

\renewcommand\vec{\bm}

\DeclareFontFamily{U}{calligra}{}
\DeclareFontShape{U}{calligra}{m}{n}{<->callig15}{}

\newcommand{\calE}{{\!\!\text{\usefont{U}{calligra}{m}{n}E}\,\,}}

\begin{document}

%\preprint

\title{Ultracold Neutrons in the Low Curvature Limit:\\
Remarks on the post-Newtonian effects}
%%Very Special Relativity and the Dirac Equation in a homogeneous Magnetic Field:
%Corrections to the Gyromagnetic Facto
%\thanks{A footnote to the article title}%

\author{Benjamin Koch}
 \email{benjamin.koch@tuwien.ac.at}
% \affiliation{Institute for Theoretical Physics, TU Wien, Wiedner Hauptstr. 8, A-1040 Vienna, Austria}
\affiliation{Institut f\"ur Theoretische Physik and Atominstitut,
 Technische Universit\"at Wien,
 Wiedner Hauptstrasse 8--10,
 A-1040 Vienna, Austria}
\affiliation{Facultad de F\'isica, Pontificia Universidad Cat\'olica de Chile, Vicu\~{n}a Mackenna 4860, Santiago, Chile}
 
\author{Enrique Mu\~{n}oz}
\email{munozt@fis.puc.cl}
\affiliation{Facultad de F\'isica, Pontificia Universidad Cat\'olica de Chile, Vicu\~{n}a Mackenna 4860, Santiago, Chile}
% \altaffiliation[Also at ]{Physics Department, XYZ University.}%Lines break automatically or can be forced with \\

\author{Alessandro Santoni}
\email{asantoni@uc.cl}
\affiliation{Institut f\"ur Theoretische Physik and Atominstitut,
 Technische Universit\"at Wien,
 Wiedner Hauptstrasse 8--10,
 A-1040 Vienna, Austria}
 \affiliation{Facultad de F\'isica, Pontificia Universidad Cat\'olica de Chile, Vicu\~{n}a Mackenna 4860, Santiago, Chile}

%\collaboration{MUSO Collaboration}%\noaffiliation

\date{\today}% It is always \today, today,
             %  but any date may be explicitly specified

\begin{abstract}
Ultracold neutrons are great experimental tools to explore the gravitational interaction in the regime of quantized states. 
From a theoretical perspective, starting from a Dirac equation in curved spacetime, we applied a perturbative scheme to systematically derive the non-relativistic Schr\"odinger equation that governs the evolution of the neutron's wave function in the Earth's gravitational field. At the lowest order, this procedure reproduces a Schr\"odinger system affected by a linear Newtonian potential, but corrections due to both curvature and relativistic effects are present. Here, we argue that one should be very careful when going one step further in the perturbative expansion. Proceeding methodically with the help of the Foldy-Wouthuysen transformation and a formal post-Newtonian $\nicefrac{1}{c^2}-$expansion, we derive the non-relativistic Hamiltonian for a generic static spacetime. By employing Fermi coordinates within this framework, we calculate the next-to-leading order corrections to the neutron's energy spectrum. Finally, we evaluate them for typical experimental configurations, such as that of qBOUNCE, and note that, while the current precision for observations of ultracold neutrons may not yet enable to probe them, they could still be relevant in the future or in alternative circumstances. 
\end{abstract}

%\keywords{Suggested keywords}%Use showkeys class option if keyword
%display desired
\maketitle

% TO-DO LIST:
% --- subsection IV.E 
% --- citations and bibliography
% --- appendices
% --- conclusions
% --- checking everything
% --- complete the outline in the Intro

%\tableofcontents

\section{Introduction}

In the last decades, there has been a surging interest in a wide variety of small-size and table-top experiments exploring the fundamental properties of the gravitational interaction: starting from optical \cite{brodutch2015post} and atom interferometry \cite{muller2008atom,ufrecht2020atom,tino2021testing}, also with the inclusion of optical lattices \cite{boada2011dirac,panda2023measuring}, getting to more exotic ideas, like Bose-Einstein condensate \cite{sekh2017bouncing}, Geonium atoms \cite{ulbricht2019gravitational,ito2021inertial} and eventual Gravitational Waves detectors \cite{tino2007possible}. \\
A very interesting possibility is offered by ultracold neutrons (UCN) \cite{ignatovich1986physics}, particles with such low energies and velocities that their wavelength become larger than typical atomic interspacing, and can therefore be stored much more easily, since they get totally reflected by many materials. Recently, UCNs
have also been employed to investigate the quantum nature of gravitational interaction. Remarkably, experiments such as qBOUNCE \cite{abele2011qbounce,jenke2009q} and GRANIT \cite{baessler2011granit,clement2022manipulation} have successfully observed gravitationally induced quantum states \cite{nesvizhevsky2010near,nesvizhevsky2003measurement, cronenberg2015gravity}. As a result, experiment involving UCN are becoming a standard option to probe fundamental physics \cite{sponar2021tests,ivanov2019probing,escobar2022testing}, in particular extensions of General Relativity (GR), like beyond-Riemannian models \cite{kostelecky2021searches,ivanov2021quantum}, Torsion contributions \cite{ivanov2015nonrelativistic}, emergent gravity proposals \cite{PhysRevResearch.3.033065,Sung:2023jir,PhysRevD.83.021502}, and much more. \\
Within this framework, GR or Standard Model extensions are usually described starting from a generalized Dirac equation, depending on the theory under analysis, embedded in the curved spacetime sourced by the Earth. From there, one can obtain the corresponding non-relativistic Hamiltonian by taking the low-curvature and low-velocity limit. Usually, the final result of this procedure can be splitted into the GR contribution and additional terms which parametrize the extension under consideration. \\
Contrary to the prevailing trend, our paper uniquely concentrates on precisely the GR contribution. At leading order, one expects to recover the Schr\"odinger equation describing a particle in the Earth's Newtonian potential, which is the theoretical picture considered in the interpretation of gravitational experiments with UCNs \cite{suda2022spectra}. However, when going deeper into the perturbative expansion, new terms get to influence the quantum dynamics and since experiments like qBounce have already reached the stunning sensibility of $10^{-17}\, eV$, one should start asking oneself to which extent will the trivial Newtonian picture hold on. Then, one of the more pragmatic purposes of this work is exactly to quantify the next-to-leading order corrections to UCNs energy spectrum in the gravitational field of the Earth. To complete this task we will have to go through several technical steps which, despite the amount of work already present in the literature, remain non-trivial. For example, the aforementioned corrections get typically sorted out by the powers of the inverse mass $\nicefrac{1}{m}$ of the fermion. Nevertheless, if not carefully considered, this choice can result in inconsistencies when dealing with the gravitational interaction (whose source is the mass itself)\cite{obukhov2002gravitational}, while still being perfectly fine in the electromagnetic sector. Also for this reason, we decided to adopt for our work the post-Newtonian approach, in which perturbations are categorized by their inverse $c^2$ power. More details on this are spread throughout the paper.\\
Therefore, the outline of this work is as follows: After a small summary illustrating our conventions, in Section II, we start from the Dirac equation in curved spacetime to evaluate the respective Hamiltonian for a static spacetime and, then, we take the low-curvature limit and introduce the post-Newtonian expansion. In Section III, we exploit the Foldy-Wouthuysen transformation to perform the non-relativistic limit, while in Section IV, with the help of Fermi coordinates, we take the perspective of an accelerated laboratory frame on the surface of the Earth. Finally, in Section V we derive the next-to-leading order corrections to the neutron's spectrum and determine their magnitude for current experiments, like qBounce. To avoid cluttering in the main text with too many calculations, we include some details on the more lengthy ones in the Appendix of this work.

\subsection{Notation and Conventions}
Before we get into the calculations, let's set up the notational conventions used along the paper: hereon, we will use greek and latin characters to label respectively spacetime and tangent space indices. As usual, the tetrad field $e^{\, a}_{\;\;\mu}$ will be used to ``translate" spacetime indices into tangent space indices and viceversa. Let's also note that to avoid confusion with the tetrads and their inverse, which are the only objects that intrinsically mix the two types of indices, we will place the (upper or lower) tangent space index as the first one appearing from left to right. Finally, time components will be indicated with $t$ for spacetime indices and with “$0$” in the tangent space, while spatial components will be differentiated by using capital letters $\{I,J,K,...\}$ for tangent space indices and lower case letters $\{i,j,k,...\}$ for the corresponding spacetime indices. \\
Since our final aim is to pursue a non-relativistic expansion within the approach of the post-Newtonian approximation \cite{chandrasekhar1965post}, we will not work in natural units to still be able to keep track of powers of $c$.  \\
For the Dirac matrices $\gamma^a$ in flat spacetime we choose the standard representation
\begin{eqnarray}
\gamma^0 = \left( \begin{array}{cc}\mathbf{1} & 0\\0 & -\mathbf{1} \end{array}\right),\,\,\gamma^I = \left( \begin{array}{cc}0 & \sigma^I\\-\sigma^I & 0 \end{array}\right) \,,
\label{eq_gamma}
\end{eqnarray}
with $\mathbf{1}$ representing a two by two identity matrix and $\vec \sigma$ the usual Pauli's matrices. From those we can define their curved spacetime version
\begin{equation}
    \underline \gamma^\mu \equiv e_{a}^{\;\;\mu} \, \gamma^{ a} \, ,
\end{equation}
which satisfy the consistent curved spacetime Clifford Algebra 
\begin{equation}
    \{ \underline \gamma^\mu, \underline \gamma^\nu \}= 2 \, g^{\mu\nu} \, .
\end{equation}
We further introduce $\sigma^{ab}$ and $\Sigma-$matrices
\begin{equation}
     \sigma^{ a b} = \frac{1}{4} [\gamma^{ a} , \gamma^ { b}] \,\,\, ,\,\,\,
     \Sigma^I = \left( \begin{array}{cc} \;  \sigma^I & 0 \\ 0 & \;\sigma^I \end{array}\right) \, .
\end{equation}
Finally, for the Minkowski metric we pick the mostly-minus convention $\eta_{\mu\nu} = diag \{ 1,-1,-1,-1\}$.

\section{Dirac Hamiltonian in Curved Spacetime}

Let us start our analysis from the Dirac equation in curved spacetime \cite{de1962representations, pollock2010dirac}
\begin{equation} \label{curvdirac} 
    (i \hbar \, \underline {\gamma}^\mu D_\mu - m c) \psi =0 \,,
\end{equation}
with $D_\mu = \partial_\mu + \Gamma_\mu$ representing the spinor covariant derivative and the Spin-Connection $\Gamma_\mu$ \cite{weinberg1972gravitation}. 
The latter is expressed through the inverse tetrads $e_{a}^{\;\;\mu}$ as
\begin{equation}\label{gammadef}
    \Gamma_\mu = \frac12 \sigma^{ a b}\, e_{a}^{\;\;\nu} \, \nabla_\mu e_{ b \nu} = \frac12 \sigma^{ a  b} \, g_{\nu\rho} \, e_{ a}^{\;\;\nu} \, \nabla_\mu e_{ b}^{\;\;\rho} \,,
\end{equation}
where
\begin{equation}\label{dercovgr}
    \nabla_\mu e_{ b}^{\;\;\rho} = \partial_\mu e_{ b}^{\;\;\rho} +\{^{\;\, \rho}_{ \mu\, \alpha}\} \, e_{ b}^{\;\;\alpha}
\end{equation}
is the usual GR covariant derivative constructed from the Christoffel Symbols $\{^{\;\, \rho}_{ \mu\, \alpha}\}$
\begin{equation}
    \{^{\;\, \rho}_{ \mu \,\, \alpha}\} = \frac{1}{2} g^{\rho \beta} ( \partial_\mu g_{\alpha \beta} +\partial_\alpha g_{\mu \beta} -\partial_\beta g_{\mu \alpha})\,.
\end{equation}
Multiplying Eq.\eqref{curvdirac} by $(g^{tt})^{-1} \underline \gamma^t$, we can manipulate it to obtain a time-evolution equation
\begin{equation}
     \mathcal{H}_D \psi  = i \hbar \frac{\partial \psi}{\partial t} \, ,
\end{equation}
where we used $x^t=ct $, and $\mathcal{H}_D$ is then the Dirac Hamiltonian for a generic spacetime 
\begin{equation} \label{diracH}
    \mathcal{H}_D = m c^2 (g^{tt})^{-1} \underline \gamma^t  -i \hbar c \, \Gamma_t - i \hbar c\,  (g^{tt})^{-1} \underline \gamma^t \underline \gamma^i D_i  \,,
\end{equation}
where repeated spatial indices are summed.\\
In general, the spacetime line element $ ds^2 = g_{\mu\nu} dx^\mu dx^\nu $ can be expressed as
\begin{equation} \label{staticmetric}
    ds^2 =  V^2 (c \, dt)^2 + g_{ij} dx^i dx^j \, ,
\end{equation}
where $V$ and $g_{ij}$ are functions of the spatial coordinates.
Remembering that the tetrads must satisfy the condition $g_{\mu \nu} = \eta_{ a  b} \, e^{a}_{\;\mu} e^{ b}_{\;\nu}$, we conveniently choose their expressions such that they do not mix time and spatial coordinates
\begin{equation}
    e^0_{\;\;i}= e^I _{\;\;t} = 0 \, , \,\,  e^{\;\;t}_{I}= e^{\;\;i} _{0} = 0\, ,
\end{equation}
and that
\begin{equation} \label{timetetrad}
   e^0_{\;\;t} = (e^{\;\;t}_{0}\, )^{-1} = V \,.
\end{equation}
In this way, the Dirac matrices with the indices referring to the curved spacetime coordinate read
\begin{eqnarray}
    \underline \gamma^t &=& e^{\;\;t}_{a} \, \gamma ^a= \frac{1}{V} \gamma^0 \,, \\
    \underline \gamma^i &=& e^{\;\;i}_{a} \, \gamma ^a = e^{\;\;i}_{J} \, \gamma ^J. \nonumber
\end{eqnarray}
With the above considerations, we can rewrite Eq.\eqref{diracH} for the particular case of a static spacetime \eqref{staticmetric}
\begin{eqnarray}
      \mathcal{H}_D = m c^2 V \gamma^0  -i \hbar c \, \Gamma_t - i \hbar c\, V   \gamma^0  \gamma^J e^{\;\;i}_{J}(\partial_i+\Gamma_i)  \,, \nonumber 
\end{eqnarray}
which, using the explicit expression \eqref{GTcomplete} of $\Gamma_t$ combined with the one for $\Gamma_i$ (see Appendix \ref{gammamu}), becomes
\begin{eqnarray} \label{diracH2}
      \mathcal{H}_D = m c^2 V \beta  -i \hbar c\, \alpha^J e^{\;\;i}_{J} \left ( \, V \partial_i+ \frac12 \partial_i V +V \, \Gamma_i \right )  \,. \nonumber \\
\end{eqnarray}
Here, we defined the matrices $ \beta \equiv \gamma^{0}$ and $\alpha^{i} \equiv \gamma^{0} \gamma^{i}$.
Note that, in curved space, the Hamiltonian \eqref{diracH2} is hermitian only with respect to the right scalar product measure \cite{arminjon2006post} including $ J \equiv \sqrt{- det(g_{ij})} $
\begin{equation} \label{jacobian}
    d^3 x \, J =d^3 x \, \sqrt{- det(g_{ij})}  \,.
\end{equation}
Equivalently, we can also implement hermiticity
through the following redefinitions for the spinor and Hamiltonian operator \cite{parker1980one,huang2009hermiticity,PhysRevD.90.045040}
\begin{eqnarray}
    \tilde \psi &=& J^{\frac12} \, \psi \,,\\
    \mathcal{\tilde H}_D &=& J^{\frac12} \, \mathcal H_D\, J^{-\frac12} \, . \nonumber
\end{eqnarray}
In this way, the Hamiltonian becomes
\begin{eqnarray} \label{finaltildeH}
      \mathcal{\tilde H}_D &=& m c^2 V \beta \\
      && -i \hbar c\, \alpha^J e^{\;\;i}_{J} \left ( \, V \partial_i + \frac12 \partial_i V -\frac12 V J^{-1} \partial_i J +V \, \Gamma_i \right ) \nonumber ,
\end{eqnarray}
which is now hermitian respect to the flat measure. From here on we will drop the “tilde”-notation for the sake of simplicity.

%%%%%%%%%%%%%%%%%%%%%%%%%%%%%%%%%
\subsection{Low-curvature Limit and post-Newtonian expansion}

At this point, we are ready to take the low-curvature or weak-gravity limit. The way we perform it is by realizing a formal $\nicefrac{1}{c} \,-$expansion of the geometrical objects around flat spacetime quantities \cite{ulbricht2019gravitational}, in the same fashion as post-Newtonian (PN) expansions
\begin{eqnarray} \label{pNexp1}
    v & \equiv& V-1 \sim {\mathcal{O}}(c^{-2}) \, ,\\
    h_{ij} & \equiv& g_{ij} - \eta_{ij} \sim {\mathcal{O}}(c^{-2}) \, , \nonumber\\
    \varepsilon_{J}^{\;\; i} & \equiv& e_{J}^{\;\; i} - \delta_{J}^{\;\; i} \sim {\mathcal{O}}(c^{-2})\, , \nonumber
\end{eqnarray}
with the perturbative objects defined above containing all the corrections starting from the smallest $\nicefrac{1}{c^2}-$order up to the largest one allowed by the context of the expansion. That also implies the following form of the jacobian in \eqref{jacobian}
\begin{equation} \label{jacobianlow}
    J = \sqrt{-det(\eta_{ij} +h_{ij})} \simeq 1-\frac12 h +{\mathcal{O}}(c^{-4}) \, , \,\,\, h\equiv \sum_i h_{ii}\,.
\end{equation}
Replacing definitions \eqref{pNexp1} and \eqref{jacobianlow} into the Hamiltonian \eqref{finaltildeH} and keeping everything up to order $\nicefrac{1}{c^2}$, we have
\begin{eqnarray} \label{lowcurvH1}
      \mathcal{ H}_D &\simeq& m c^2 \beta+m c^2 \beta v -i \hbar c\, \alpha^J \left ( \, (1+v) \delta_{J}^{\;\; i} + \varepsilon^{\;\;i}_{J} \, \right ) \partial_i \nonumber \\
      && -i \hbar c\, \alpha^J \delta_J^{\;\;i}\left ( \, \frac12 \partial_i v + \Gamma_i + \frac{1}{4}\partial_i h \right ) \,,
\end{eqnarray}
where each perturbation term $v$, $h$ or $\varepsilon$ is intended to be expanded up to highest possible order while keeping the Hamiltonian at order $\nicefrac{1}{c^2}$.\\
In practice, this approach is equivalent to the one used for PN calculations \cite{chandrasekhar1965post,nelson1990post}, which have demonstrated to be very powerful when dealing with gravitational systems. By virtue of this analogy, we borrow some of the PN-vocabulary to classify corrections in a convenient way: in particular, terms of order $\nicefrac{1}{c^N}$ in our scheme, will correspond to $\frac{N}{2}$PN-corrections. \\
%In this sense, $\nicefrac{1}{c^2}-$terms in \eqref{lowcurvH} represent $1PN-$correction while $\nicefrac{1}{c^2}-$terms
\begin{table}[ht]
\centering
\begin{tabular}[t]{ccc}
\toprule
$\;\;${$\nicefrac{1}{c}-$order}$\;\;$ & $\;\;$PN-equivalent$\;\;$ \\ 
    \midrule
    {$1$} & 0PN \\
    $c^{-1}$ & 0.5PN \\
    $c^{-2}$ & 1PN \\
    \bottomrule
\end{tabular} 
\caption{Correspondence table between our formal expansion and the post-Newtonian one (also look at Fig.1 in \cite{bern2019black})}.
\label{tab:PNcorresp}
\end{table}\\
In this sense, the Hamiltonian \eqref{lowcurvH1} must be interpreted as the 1PN-version of the complete expression \eqref{finaltildeH}. Thus, apart from the 0PN Newtonian contribution, it will also include 0.5PN and 1PN corrections that are the focus of this work. However, it is important to highlight that, even if here “hybrid” 0.5PN contributions are present due to the structure of the Dirac Hamiltonian, in the rest of the paper we will see that those are gonna become additional 1PN perturbations after the non-relativistic limit is taken.\\
From Eq.\eqref{lowcurvH1} we can also note that, at this stage of the calculation, the distinction between capital and lowercase spatial indices becomes irrelevant. In fact, the information on the perturbations around the flat spacetime (up to the relevant order) is already encoded in the $v-$ and $\varepsilon-$objects. \\
Therefore, with a little abuse of notation, from now on we will drop this distinction to avoid unnecessary complexities in the reading 
\begin{equation} \label{newindnot}
    \alpha^I \to \alpha^i \,\, , \,\,\, \delta_{J}^{\;\;i} \to \delta_{j}^{\;\;i} \,\, , \,\,\, \varepsilon_{J}^{\;\;i} \to \varepsilon_{j}^{\;\;i} \,.
\end{equation}
This allows us to write the Hamiltonian \eqref{lowcurvH1} in the following convenient way
\begin{eqnarray} \label{lowcurvH}
      \mathcal{ H}_D &\simeq& m c^2 \beta+m c^2 \beta v \\
      && -i \hbar c\, \alpha^i \left ( \, (1+v) \,\partial_i + \varepsilon^{\;\;j}_{i} \partial_j \, \right .  \nonumber \\
      &&  \left . \;\;\;\;\;\;\;\;\;\;\;\;\;\;\; +\frac12 \partial_i v + \Gamma_i + \frac{1}{4}\partial_i h \right ) \nonumber \,.
\end{eqnarray}
%where the $\delta$'s are always indicating identity tensors.

\subsection{Simplified expression for $\Gamma_i$}

Starting from Eq.\eqref{GIcomplete} for $\Gamma_i$, we now want to expand it and obtain its 1.5PN expression, which is relevant for Hamiltonian \eqref{lowcurvH}. Making use of the new notation \eqref{newindnot}, we have the result
\begin{eqnarray} \label{GIexpanded}
    \Gamma_i &\simeq& \frac{1}{2} \sigma^{kl} ( \eta_{mn } \delta_{k}^{\;\;m} \partial_ i \epsilon_{l}^{\;\;n } + \delta_{k}^{\;\;m} \delta_{l}^{\;\;n } \partial_n h_{im} ) \nonumber \\
    &\simeq& \frac{1}{2} \sigma^{kl} ( \partial_ i \epsilon_{k}^{\;\;l } - \partial_k h_{il} ) \, ,
\end{eqnarray}
where repeated spatial indices are summed up independently of their upper or lower position. Nevertheless, there is another simplification that can be done in a few steps: first of all, let's observe that, expanding up the tetrad conditions $e_{a}^{\;\; \mu} e^{a}_{\;\; \nu} = \delta_{\;\nu}^\mu$ and $ g_{\mu\nu} = e^{a}_{\;\; \nu} e^{a}_{\;\; \nu} \eta_{a b}$, we obtain respectively the equations
\begin{eqnarray} \label{tetradcond}
     \varepsilon^{j}_{\;\; i} &=& -\,\varepsilon_{i}^{\;\; j} \,, \\
    h_{ij} &=& - \,\varepsilon^{i}_{\;\; j} -\varepsilon^{j}_{\;\; i} =  \varepsilon_{i}^{\;\; j}+\varepsilon_{j}^{\;\; i} \nonumber \,.
\end{eqnarray}
Thus, it is straightforward to see that we can always make the following choice for the tetrads
 \begin{eqnarray}
  \varepsilon_{i}^{\;\;j} &=& -\,\varepsilon^{j}_{\;\;i}  \equiv \frac12 h_{ij} \,,
\end{eqnarray}
so that the conditions \eqref{tetradcond} are fullfilled up to the relevant $\nicefrac{1}{c^3}-$order. This way, the matrix representing the tetrad tensors will be symmetric in the sense that
\begin{equation} \label{symmtetrad}
    \varepsilon_{i}^{\;\; j} = \varepsilon_{j}^{\;\; i} \, .
\end{equation}
On the other hand, the above relation implies that contractions of the type $ \sigma^{kl} \, \varepsilon_{k}^{\;\; l} $ vanish, because of the (anti)symmetries of the involved objects. Therefore, we can finally simplify the $\Gamma_i$ to the expression
\begin{equation} \label{simplegammaj}
    \Gamma_i \simeq -\frac{1}{2} \sigma^{kl}   \partial_k h_{i l}  \,.
\end{equation}

%%%%%%%%%%%%%%%%%%%%%%%%%%%%%%%%%%%%
\section{Non-Relativistic Limit} \label{NRlimit}

We shall now proceed with the low-velocity or non-relativistic limit $|\vec p| << m c$. In order to do that, we will apply the Foldy-Wouthuysen (FW) transformation \cite{foldy1950dirac}, a well-known procedure to decouple Dirac spinors into its
positive and negative energy components.
This is usually valid up to some order in $m^{-1}$, which would otherwise get mixed by the $\alpha-$matrices. However, since we are working in the PN hierarchy of perturbative corrections, we will keep using $\nicefrac{1}{c}$ as our formal expansion parameter. That also allows us to avoid the inconsistencies raised in \cite{obukhov2002gravitational} when using the standard FW transformation in the gravitational context.

%%%%%%%%%%%%%%%%%%%%%%%%%%%%
\subsection{Foldy-Wouthuysen Transformation}

The first step in the FW approach is to divide the Hamiltonian \eqref{lowcurvH} into the ``even" (non-mixing) operator $\mathcal{E}$ and the ``odd" (mixing) operator $\Theta$ \cite{jentschura2014foldy}
\begin{equation} \label{Hfwstart}
    \mathcal{H}_D = \beta  m c^2+\mathcal{E}+ \Theta\,,
\end{equation}
with
\begin{equation}\label{epsilontheta}
    \left\{\begin{array}{l} 
    \Theta= -i \hbar c\, \alpha^i \left (  (1+v) \, \partial_i + \varepsilon^{\;\;j}_{i} \partial_j + \frac14 \partial_i (2v +h) + \Gamma_i \right )  , \\
    \, \\
    \mathcal{E}= m c^2 \beta v \, , \\ 
    \end{array} \right. 
\end{equation}
The even and odd operators satisfy the following (anti)commutation relation
\begin{equation}
  [\mathcal{E}\, , \beta]= 0 \, , \,\,\,  \{\alpha^i,\beta\} = 0\to   \{ \Theta, \beta \} = 0 \, ,
\end{equation}
and we can easily see that, due to their expressions and the relations \eqref{pNexp1}, we have at lowest order
\begin{equation} \label{EOorder}
    \mathcal{E}\sim {\mathcal{O}}(1) \,,\,\, \Theta \sim {\mathcal{O}}(c)\,.
\end{equation}
In our formalism, the typical unitary FW transformation $U=e^{iS}$, with $S$ hermitian, is defined as
\begin{equation}
    S= -i \frac{\beta }{m c^2}\Theta \sim {\mathcal{O}}(c^{-1}) \, .
\end{equation}
Then, the following expansion is valid up to 1PN order
\begin{eqnarray}\label{HFWcommut}
    \mathcal{H}_{FW} &=& e^{iS} \mathcal H_D e^{-iS} \\
    &=& \mathcal H_D + i [S,\mathcal H_D] \nonumber\\
    && + \frac{i^2}{2!} [S,[S,\mathcal H_D]] +\frac{i^3}{3!} [S,[S,[S,\mathcal H_D]]] \nonumber\\
    && +\frac{i^4}{4!} [S,[S,[S,[S,\mathcal H_D]]]+ {\mathcal{O}}(c^{-3}) \nonumber\,.
\end{eqnarray}
Starting from these settings, after three consecutive FW transformations, we end up with the Hamiltonian \cite{hehl1990inertial}
\begin{eqnarray} \label{HFWafter}
    \mathcal H_{FW} &=& \beta m c^2 +\mathcal{E} + \frac{\beta \Theta^2}{2m c^2} \\
    &&+\frac{\beta}{8 m^2 c^4} [[\Theta,\mathcal{E}],\Theta] - \frac{ \beta}{8m^3 c^6} \Theta^4 + {\mathcal{O}}(c^{-3})\nonumber \,.
\end{eqnarray}
in which the matter and anti-matter sectors have been decoupled up to order $\nicefrac{1}{c^3}$. Thus, each term in \eqref{HFWafter} must be calculated up to the relevant order for our approximation. For example, to keep only terms at most 1PN, we should compute $[[\Theta,\mathcal{E}],\Theta]$ at least to order $c^{2}$, and so on. More details on the FW transformation are included in Appendix \ref{FWmethod}.\\
At this point, it is straightforward to obtain the non-relativistic Hamiltonian $H_{NR}$ describing the fermion dynamics by simply selecting the positive energy solutions of $\mathcal{H}_{FW}$, and neglecting its constant mass term that would only produce an overall shift to the energy spectrum. After working out every single commutator in \eqref{HFWafter} and defining the shifted spatial metric correction $\tilde h_{ij}$
\begin{equation}
\tilde h_{ij} \equiv h_{ij} + v \, \delta_{ij} \,,
\end{equation}
the final result can be expressed in a surprisingly compact and practical form
\begin{eqnarray} \label{HNR}
    H_{NR} &=& m c^2 v -\frac{\hbar^4 \partial_i ^4}{8 m^3 c^2} \\
    && -\frac{\hbar^2}{2m} \left ( 
    \partial_ i ^2 +\tilde h_{ij} \partial_ i \partial _j + \partial_ i \tilde h_{ij}  \partial _j +\frac14 \partial_i \partial_j \tilde h_{ij} \nonumber \right .\\
    && \left . \;\;\;\;\;\;\;\;\;\;\;+\frac{i}{2} \epsilon^{ijk} \sigma^k \partial_i \tilde h_{jl} \partial_l + \frac{i}{4} \epsilon^{ijk} \sigma^k \partial_i \partial_l \tilde h _{jl} 
    \right) \nonumber \, ,
\end{eqnarray}
with $\epsilon^{ijk}$ being the Levi-Civita symbol. For the consistency of the expression, $v$ must be calculated here up to order $\nicefrac{1}{c^{4}}$ while $\tilde h_{ij}$ up to $\nicefrac{1}{c^{2}}$. This is one of the main results of this paper due to its compactness and general validity for static weak gravity scenarios. We summarize again the considerations used to achieve it
\begin{itemize}
    \item $g_{0i}=0$ \eqref{staticmetric},
    \item $v \,,\, h_{ij} \,, \,\varepsilon_{j}^{\;\;i} \sim {\mathcal{O}}(c^{-2})$ \eqref{pNexp1},
    \item $\varepsilon_{i}^{\;\;j}=\varepsilon_{j}^{\;\;i}$ \eqref{symmtetrad}.
\end{itemize}

\subsection{Examples and Comparison}

In this section, we would like to consider a few applications for Eq.~\eqref{HNR} and some comparisons with the literature. 

\subsubsection{Diagonal Spacetime Metrics}

Let us start by considering the special case of a diagonal static metric with the form
\begin{equation}
    g_{\mu\nu} = diag \{ V^2, -W^2, -W^2, -W^2 \} \, .
\end{equation}
When assuming the weak gravity limit in \eqref{pNexp1}, we define the additional perturbative quantity
\begin{equation}
    w \equiv W-1 \sim \mathcal{O}(c^{-2}) \, .
\end{equation}
Therefore, it will be sufficient to replace $h_{ij} = -2 w \, \delta_{ij}$ in Eq.~\eqref{HNR} to obtain the relevant Hamiltonian expression for this case, that becomes
\begin{eqnarray} \label{HNRdiagonal}
    H_{NR} &=& m c^2 v -\frac{\hbar^4 \partial_i ^4}{8 m^3 c^2} \\
    && -\frac{\hbar^2}{2m} \left ( 
     (1+v-2w) \partial_ i^2 + \partial_ i (v-2w)  \partial _i \right . \nonumber \\
    && \;\;\;\;\;\;\;\;\left. +\frac14 \partial_i^2 (v-2w) \nonumber +\frac{i}{2} \epsilon^{ijk} \sigma^k \partial_i (v-2w) \partial_j \right )  \, ,
\end{eqnarray}
which corresponds to the result in \cite{silenko2005semiclassical}.

\subsubsection{Schwarzschild Metric} 

A straightforward application of the previous formulae is for the Schwarzschild spacetime. In fact, considering the low-curvature limit in isotropic coordinates, we obtain the following spacetime element
\begin{equation} \label{schwar}
    ds^2= \left(1+\frac{2\Phi_S}{c^2}+\frac{2\, \Phi_S^2}{c^4}\right)(c \,dt)^2-\left(1-\frac{2 \Phi_S}{c^2}\right) \, \vec {dx} ^2 \,,
\end{equation}
with $\Phi_S \equiv- \frac{GM}{r}$ the gravitational potential external to the spherical mass source $M$, and $r = \sqrt{x^2+y^2+z^2}$ the coordinate distance from its center. In the temporal and spatial components of the metric, we have kept terms up to the relevant order for our case. This expression leads us to 
\begin{eqnarray}
    v = \frac{\Phi_S}{c^2}+\frac{ \Phi_S^2}{2 c^4} \, , \,\,\, w = -\frac{\Phi_S}{c^2} \, ,
\end{eqnarray}
which replaced into \eqref{HNRdiagonal} gives back the non relativistic Hamiltonian for the Schwarzschild metric
\begin{eqnarray} \label{Hschw}
    H_{S} &=& m \Phi_S \left (1 +\frac{\Phi_S}{2c^2} \right ) -\frac{\hbar^2}{2m} \left (1+ \frac{ 3\Phi_S}{c^2} \right ) \partial^2_i -\frac{\hbar^4 \partial_i^4}{ 8 m^3c^2} \nonumber \\
    && + \frac{3 \hbar^2}{8m c^2} \vec \partial \cdot \vec g + \frac{3 \hbar^2}{2m c^2} \, \vec g \cdot \vec \partial +\frac{3 i\hbar^2}{4mc^2} \vec \sigma \cdot (\vec g \times \vec \partial) \nonumber \,,\\
\end{eqnarray}
where we defined the “Newtonian” gravitational acceleration vector $\vec g$ as
\begin{equation} \label{gdef}
    g^i \equiv  - \partial_i \Phi_S \, .
\end{equation}
Expression \eqref{Hschw} matches the results in \cite{jentschura2014foldy, jentschura2013nonrelativistic, PhysRevD.90.045040}, and also \cite{fischbach1981general} if we neglect the Darwin term $\propto \vec p \cdot \vec g \sim \vec \partial ^2 \Phi_S $, which outside the source of the gravitational field does not matter anyway.

\subsubsection{Eddington-Robertson Metric}

Finally, we want to discuss here a particular case that will also come in handy later in the paper: the Eddington–Robertson (ER) parametrized post-Newtonian metric \cite{giulini2023coupling, lammerzahl1995hamilton} 
\begin{equation} \label{ERmetric}
    ds^2 = \left(1+\frac{2\Phi_N}{c^2}+\frac{2 \beta \, \Phi^2_N}{c^4}\right)(c \,dt)^2-\left(1-\frac{2 \gamma \Phi_N}{c^2}\right) \, \vec {dx} ^2 \,,
\end{equation}
where $\Phi_N$ is the usual Newtonian potential for an extended classical body with density $\rho(\vec x)$
\begin{equation}
    \Phi_N (\vec x)= G \int _{\text{Source}} d^3x' \frac{ \rho(\vec x')}{\norm{\vec x- \vec x'}} \, ,
\end{equation}
while the parameters $\beta$ and $\gamma$ account for possible deviations from GR (in which $\beta=\gamma=1$), and should not be mistaken with the Dirac matrices above-presented. We re-label these parameters respectively as $b$ (for $\beta$) and $d$ (for $\gamma$) to avoid confusion. The GR limit of \eqref{ERmetric} is the solution of Einstein equation in a $\nicefrac{1}{c}-$expansion for a static source. The ER metric is the simplest example of a metric in the general parameterized post-Newtonian (PPN) formalism, which provides a general framework for testing metric theories of gravity in the weak-field regime. For a more exhaustive discussion on this topic, see \cite{poisson2014gravity}. \\
Considering a spherically symmetric source and limiting ourselves to its exterior, $\Phi_N$ reduces to the Schwarzschild expression $\Phi_S=-\frac{GM}{r}$, yielding what we call the ER-Schwarzschild (ERS) metric 
\begin{equation}\label{ERSmetric}
    ds^2= \left(1+\frac{2\Phi_S}{c^2}+\frac{2b \, \Phi_S^2}{c^4}\right)(c \,dt)^2-\left(1-\frac{2 d \Phi_S}{c^2}\right) \, \vec {dx} ^2 \,, \\
\end{equation}
which is trivially equivalent to \eqref{schwar} when $b=d=1$ as expected. Following the same procedure as before, the ERS Hamiltonian reads
\begin{eqnarray} \label{ERShamil}
    H_{ERS} &=& m\Phi_S \left (1 +(2b-1)\frac{\Phi_S}{2c^2} \right ) \\
    && -\frac{\hbar^2}{2m} \left (1+(1+2d) \frac{ \Phi_S}{c^2} \right ) \partial^2_i -\frac{\hbar^4 \partial_i^4}{ 8 m^3c^2} \nonumber \\
    && + (1+2d)\frac{\hbar^2}{8m c^2} \vec \partial \cdot \vec g + (1+2d)\frac{\hbar^2}{2m c^2} \, \vec g \cdot \vec \partial \nonumber\\
    && +(1+2d)\frac{i\hbar^2}{4mc^2} \vec \sigma \cdot (\vec g \times \vec \partial) \nonumber \, ,
\end{eqnarray}
from which we recover again \eqref{Hschw} if $b=d=1$.

\section{Proper Laboratory Frame}

Our final goal is to apply our derivation to experiments and observations made in small laboratories on Earth's surface. For this purpose, (almost) global coordinates $\{ x^\mu \} $ like the Schwarschild ones are clearly not the most suitable option, since they would lead to difficulties with the definitions of time intervals and physical distances. \\
The most natural possibility, instead, is to work in the proper reference frame of the laboratory, which can be done exploiting Fermi coordinates (FC) \cite{o1957fermi,manasse1963fermi,synge1960relativity} extended to the case of accelerated motion \cite{li1979coupled} thanks to Fermi-Walker transport \cite{alsing2009spin}. In fact, the laboratory does not follow a geodesic motion, since it is accelerated upwards by the normal force exerted by the Earth's surface itself. Therefore, this should be the natural framework when one is interested into local observations. This approach also has the advantage to make coordinate time and lengths coincides with their corresponding physical quantities, avoiding any possible confusion and need for rescalings.

\subsection{Fermi Coordinates}

The main philosophy of FC is to approximate a small enough region of spacetime around the worldline $\xi^\mu (x)$ of an observer. This task is achieved by considering the observer's proper time $\tau$ and constructing an Euclidean grid $\{X^i\}$ comoving with the observer. For further details on the geometric construction of FC see \cite{maluf2008construction,klein2008general} and references therein. In the following, objects evaluated on the observer's worldline (i.e: $X^i = 0$) are denoted when possible by a bar over them: $O|_\xi = \bar O$. Furthermore, since here we are particularly interested in static observers, we are free to align the $Z-$direction with the local acceleration $\vec a$ along all the path of the worldline. \\
The consistency of this treatment is governed by a new small parameter to be introduced in the picture, $\norm{X} / \mathcal R<<1$, with $\norm{X}$ representing the typical length scale of the experiment and $\mathcal R$ symbolically defined by \cite{audretsch1994ramsey, alibabaei2023geometric}
\begin{equation}
    \mathcal R = 
    \min \left ( \, {{\norm{R_{\mu\nu\rho\sigma}}}^{-\frac12}} \,,\,\,  \frac{ \norm{R_{\mu\nu\rho\sigma}}}{\norm{R_{\mu\nu\rho\sigma ; \alpha }}} \, , \,\,\frac{c^2}{\norm{\vec a}} \right )  \, , 
\end{equation}
where we used the semicolon in the Riemann tensor $R_{\mu\nu\rho\sigma ; \alpha
}$ to indicate its covariant derivative.

\subsection{Metric in Fermi Coordinates}

When mapped to our notational conventions, the general form of the Fermi metric in the proper reference frame experiencing an acceleration $\vec a$ is \cite{alibabaei2023geometric}
\begin{eqnarray} \label{fermimetric1}
    g^F_{\tau\tau} &=& \left (1+ \frac{\vec a \cdot \vec X}{c^2} \right )^2 - \bar R^{F}_{tltm} X^l X^m +{\mathcal{O}}(\,\norm{X}^3) \,, \nonumber\\
    g^F_{\tau i} &=& -\frac23 \bar R^F_{tlim} X^l X^m +{\mathcal{O}}(\,\norm{X}^3)\, \\
    g^F_{ij} &=& -\delta_{ij} -\frac{1}{3} \bar R^{F}_{ijlm} X^l X^m +{\mathcal{O}}(\,\norm{X}^3)\, , \nonumber
\end{eqnarray}
where $\bar R^{F}_{\mu\nu\alpha\beta} $ is the Riemann tensor in FC evaluated on the observer's worldline. Its Fermi expression, due to gauge covariance, can be evaluated starting from the Riemann tensor in some prior coordinates
\begin{eqnarray} \label{Rtransform}
    \bar R^{F}_{\mu\nu\alpha\beta} &=&  \bar R_{\rho\sigma \kappa \gamma} \,  \bar \Lambda^{\;\,\rho}_{\mu} \, \bar \Lambda^{\;\,\sigma}_{\nu} \,\bar \Lambda^{\;\,\kappa}_{\alpha} \, \bar \Lambda^{\;\,\gamma}_{\beta} \, , \\
    &\text{with}& \;\; R_{\mu\nu\alpha\beta} = g_{\mu \rho} R^{\rho}_{\,\,\nu\alpha\beta} \nonumber \,,
\end{eqnarray}
where $\bar \Lambda$ represents the coordinate transformation matrix evaluated on the worldline and, by construction \cite{synge1960relativity}, it coincides with the tetrad matrices. \\
Clearly, in our weak-gravity framework, the Riemann tensor $R_{\mu\nu\alpha\beta}$ should be treated as an ${\mathcal{O}}(c^{-2})-$object, since its leading contributions are at least linear in $h_{ij}$ \cite{marzlin1994fermi}. Note also that, working in a static context, we will have $\bar R_{t l i m }=0$ implying $g^F_{\tau i}=0$, as expected. There exists in the literature a more general metric than \eqref{fermimetric1} in which the effects of laboratory's rotation are also taken into account \cite{thorne2000gravitation}. However, in this work we shall limit ourselves to \eqref{fermimetric1} for consistency.\\
Therefore, the relevant quantities to consider for the calculation of \eqref{HNR} are
\begin{eqnarray} \label{vhfermi}
    v^F &=& \frac{\vec a\cdot \vec X}{c^2} - \frac12 \bar R^F_{\tau l \tau m} X^l X^m \,,\\
    h^F_{ij} &=& -\frac{1}{3} \bar R^F_{iljm} X^l X^m \nonumber \,.
\end{eqnarray}
Here $\bar R^F_{\tau l \tau m}$ has to be calculated up to order $\nicefrac{1}{c^4}$, starting from equation \eqref{Rtransform}
\begin{equation}
   \bar R^F_{\tau l \tau m} X^l X^m = (1+ 2 \,\bar v) \bar R_{tltm} X^l X^m  + 2 \, \bar \varepsilon_{l}^{\;\;k} \bar R_{tktm} X^l X^m  ,
\end{equation}
while $\bar R^F_{iljm}$, at order $\nicefrac{1}{c^2}$, just coincides with its expression in the prior coordinates system 
\begin{equation}
    \bar R^F_{iljm} \simeq \bar R_{iljm} \, .
\end{equation}

\subsection{Fermi Hamiltonian}

Replacing \eqref{vhfermi} in the Hamiltonian \eqref{HNR}, we can split it into the sum of two contributions
\begin{eqnarray} \label{HNRsum}
    H_{NR} = H_N + H_{NLO},
\end{eqnarray}
with the Newtonian Hamiltonian $H_N$ defined as
\begin{equation} \label{HNexpr}
     H_N = m \, \vec a \cdot \vec X -\frac{\hbar^2}{2m} 
    \partial_ i ^2 = m a Z -\frac{\hbar^2}{2m} 
    \partial_ i ^2 \,,
\end{equation}
while the next-to-leading order (NLO) part, containing all 1PN and $O(\,\norm{X}^2)$ corrections, is
\begin{eqnarray} \label{finalHNLO}
    H_{NLO} &=& -\frac{mc^2}{2} \bar R^F_{tltm} X^l X^m -\frac{\hbar^4 \partial_i ^4}{8 m^3 c^2} \\
    && -\frac{\hbar^2}{2m} \left ( \tilde h^F_{ij} \partial_ i \partial _j + \partial_ i \tilde h^F_{ij}  \partial _j +\frac14 \partial_i \partial_j \tilde h^F_{ij} \nonumber \right .\\
    && \left . \;\;\;\;\;\;\;\;\;\;\;+\frac{i}{2} \epsilon^{ijk} \sigma^k \partial_i \tilde h^F_{jl} \partial_l + \frac{i}{4} \epsilon^{ijk} \sigma^k \partial_i \partial_l \tilde h^F _{jl} 
    \right) \nonumber \, .
\end{eqnarray}
To make this picture coherent with our formulation, we assume that $\vec a$ does not depend on time, or that its dependence is sufficiently weak to be negligible over the relevant time scale involved in the physical process we want to study.\\
As clearly shown above \eqref{vhfermi}, to proceed with the calculations we also have to make some assumptions about spacetime geometry and its “prior" coordinates. To avoid additional complications at this stage, in the following we will model the Earth as a sphere of radius $R$, which naturally leads us to the choice of the ERS metric \eqref{ERSmetric}. This decision also allows us to look after eventual GR departures.\\
Note that, as far as static observers on the Earth's surface are concerned, the spatial components of their quadrivelocity $U^\alpha$ are zero. Thus, their spatial position will be constant and simply set to
\begin{equation} \label{staticworldline}
    \xi^i(x) = (0,0,R) \,,
\end{equation}
due to our reference frame choices. The full expression for the NLO Hamiltonian in the ERS spacetime together with other details are included in Appendix \ref{HERSinFermi}. \\

%%%%%%%%%%%%%%%%%%%%

\subsection{Theoretical Local Acceleration value}

In the next sections, we will consider the local acceleration just as a parameter that is determined experimentally. Nevertheless, for completeness, we would like to include here some remarks on the theoretical values predicted for the experienced acceleration. Working with a static observer, we know its spacetime acceleration $a^\mu = U^\alpha \nabla_\alpha U ^\mu$ along $\xi^\mu$, when translated in the mostly-minus convention, must be given by \cite{carroll2019spacetime,dahia2011static}
\begin{equation}
    a_\mu  = -c^2 \,\nabla_\mu \ln V |_\xi  =  -\frac{c^2}{1+\bar v} \, \partial_\mu v |_\xi \, .
\end{equation}
In our perturbative scheme, considering the ERS metric \eqref{ERSmetric} with $v = \frac{\Phi_S}{c^2} +\frac{2b-1}{2}\frac{\Phi_S^2}{c^4}$ and keeping everything up to 1PN order, we have
\begin{eqnarray}
    a_\mu  &=& -\frac{c^2}{1+\frac{\bar \Phi_S}{c^2} +\frac{2b-1}{2}\frac{\bar \Phi_S^2}{c^4}} \,\partial_\mu \left (\frac{\Phi_S}{c^2} +\frac{2b-1}{2}\frac{ \Phi_S^2  }{c^4} \right )\bigg |_\xi\nonumber \\
    & \simeq& -(1+2(b-1)\frac{\bar \Phi_S}{c^2}) \, \partial_\mu \Phi_S |_\xi  \,,
\end{eqnarray}
from which we readily see that $a_t= 0$. Therefore, raising the acceleration index with the metric, we get
\begin{eqnarray}
    a^t &=& \bar g^{t\mu} a_\mu = 0 \,, \\
    a^i &=& \bar g^{i\mu} a_\mu =  -(1+ \frac{2d \bar \Phi_S}{c^2}) \, a_i \\
    &=&  (1+2(b+d-1)\frac{\bar \Phi_S}{c^2}) \, \partial_i \Phi_S |_\xi  \, .\nonumber
\end{eqnarray}
To calculate the corresponding value on the Fermi frame we just exploit the coordinate transformation $\Lambda$
\begin{eqnarray}
    && (a^F)^\tau = \bar \Lambda^{\tau}_{\;\;\nu} \, a^\nu = \bar \Lambda^{\tau}_{\;\;t} \, a^t = 0 \, , \\
    && (a^F)^i = \bar \Lambda^i_{\;\;\nu} \, a^\nu = a^i+ \bar \varepsilon^{\,i}_{\;\;j}  \,a^j = (1- \frac{d\,\bar \Phi_S}{c^2} ) \, a^i \, .\nonumber
\end{eqnarray}
Finally, remembering the definition \eqref{gdef} of the gravitational acceleration vector, the effective acceleration experienced by the Fermi observer will simply be
\begin{equation}
    (a^F)^i = -(1+(2b+d-2)\frac{\bar \Phi_S}{c^2}) \, \bar g^i \,, 
\end{equation}
which, as expected, coincides with the Newtonian result at leading order and is in agreement with the equivalence principle by construction. Clearly, in real-life experiments, there are other effects influencing the effective acceleration value, like the Earth's rotation.\\
Note that the above result for the acceleration can also be obtained starting directly from the ERS Hamiltonian \eqref{ERShamil}, and performing the Fermi transformation $\Lambda$ a posteriori: in fact, after the transformation, it is sufficient to identify the “effective acceleration” as the coupling to the term linear in $Z$ and proportional to $m$. Embracing this philosophy, we also have to remember that coordinate time in this case does not coincide with the proper one, leading to a rescaling also of the Hamiltonian itself $H^{(\tau)} = \frac{1}{1+\bar v} H^{(t)}$. The equivalence of this approach points towards the fact that the order in which one performs the FW transformation and the change to Fermi coordinates, at the end of the day, does not affect relevant quantities, as one could also expect. However, we will not include more details on that since it goes beyond the scope of this paper.

\subsection{Remarks on higher in $\norm{X}-$orders}
An analysis close to ours is carried out in \cite{alibabaei2023geometric}, where the authors argue that strictly speaking, if one would also like to formally expand the Hamiltonian in $\norm{X}$, then the derivatives should be considered as order $\sim \nicefrac{1}{\norm{X}}$ implying the need to take into account orders higher than $\norm{X}^2$ in \eqref{fermimetric1}. However, here we adopt another view for the Fermi expansion: by momentarily adopting dimensionless quantities, let's imagine to divide the Hamiltonian \eqref{finalHNLO} by $m c^2$. Defining the momentum operator
\begin{equation}
    p^i \equiv - i \hbar \partial_i \, ,
\end{equation}
we can then think for all terms enclosed by round brackets in \eqref{finalHNLO} to be of order $\sim (\frac{p}{mc})^2 (1+ \tilde h^F)$, where we remember that $\tilde h^F_{ij}$ is already dimensionless. Thus, it is clear that all terms ${\mathcal{O}}(\,\norm{X}^3)$ we could consider in the Fermi metric expansion would give smaller and smaller corrections. Thus, holding onto to the principle of “least action”, here meaning smallest modification with largest consequences, we stick with perturbations at most $\sim \norm{X}^2$ in $g^F_{\mu\nu}$.

%%%%%%%%%%%%%%%%%%%%%%%%
\section{Application to UCN and qBounce}

In this section, we would like to quantitatively study the effects of the NLO corrections within an experimental setup analogous to the one of qBounce \cite{abele2011qbounce,Sedmik:2019twj,suda2022spectra}.
Values for kinematic parameters and constants in this configuration are summarized in table \ref{tab:qBounce}.  
\begin{table}[ht]
\centering
\begin{tabular}[t]{lcc}
\toprule
{qBounce Parameters} & Values \\ \midrule
    {Neutron mass $m$} & $1.675\times 10^{-27} \,kg$  \\
    {Earth mass $M$} & $ 5.9726 \times 10^{24} \,kg$ \\
    {Newton constant $G$}  & $6.6743\times10^{-11} \, \frac{N \cdot m^2}{kg^2}$\\ \midrule
    {Local acceleration $a$}  & $9.8049 \, \nicefrac{m}{s^2}$ \\
    {Longitudinal velocity $\upsilon_\perp$}  & $\sim 4-10 \, \nicefrac{m}{s}$ \\
    {Vertical velocity $\upsilon^Z$}  & $\sim 10 \, \nicefrac{cm}{s}$ \\
\bottomrule
\end{tabular} 
\caption{Some constants and parameters values used to fit the qBounce experiment \cite{abele2011qbounce}. The local acceleration value here, is determined through a falling corner cube classical experiment.}
\label{tab:qBounce}
\end{table}

\subsection{Decoupling the $XY-$Dynamics}

At leading order, neutrons are just affected by a $Z-$dependent potential given in \eqref{HNexpr}. This implies that their dynamics can be factorized into a longitudinal $XY-$component and a transverse $Z-$component, which is reflected in their wavefunction as
\begin{equation}
    \Psi(\vec X) = \phi(X,Y) \, \varphi(Z) \, .
\end{equation}
The free $XY-$motion in these experiments can be well described by semiclassical laws, considering the UCN's longitudinal states as normalized wave-packets centered on their classical trajectory \cite{martin2018testing,escobar2022testing}. For our purposes, it can be therefore modeled as
\begin{equation}
   \phi(\vec X_\perp, \tau)\equiv \frac{1}{\sqrt{\pi}\sigma} e^{\frac{i}{\hbar} \vec k_\perp \cdot \vec X_\perp - \frac{(\vec X_\perp -\vec X^{cl}_\perp)^2}{2\sigma^2}} \,,
\end{equation}
with $\vec X_\perp  = (X,Y)$ and
\begin{equation} 
\begin{cases}
\vec X_\perp^{cl} (\tau ) = (X_{cl}(\tau),Y_{cl}(\tau)) \,, \\
\vec k_\perp = (k^x, k^y) \, .
\end{cases} 
\end{equation}
respectively indicating the classical horizontal coordinates and momenta of the UCNs. \\
Later on, we will be mainly interested in the energy spectrum. Then, for us it is sufficient to consider $Z-$energy eigenstates and thus set the $XY-$initial state in the origin of the laboratory frame
\begin{equation} \label{initialstate}
   \phi(\vec X_\perp) \equiv \phi(\vec X_\perp, \tau=0)= \frac{1}{\sqrt{\pi}\sigma} e^{\frac{i}{\hbar} \vec k_\perp \cdot \vec X_\perp - \frac{\vec X_\perp ^2}{2\sigma^2}} \,.
\end{equation}
In the following we want to study the details of the fully-quantum transverse dynamics, by integrating out the horizontal degrees of freedom. Doing this, we can derive an effective one-dimensional Hamiltonian guiding the vertical $Z-$evolution
\begin{equation} \label{Hzform}
    H^{(Z)} = \int d^2 X_\perp \phi^*(\vec X_\perp) \, H \,  \phi(\vec X_\perp) \,.
\end{equation}
Note that the spatial spreading $\Delta X = \Delta Y = \sigma / \sqrt{2}$ is not directly known for the UCNs in the qBounce experiment. Nevertheless, we know for sure it has to be bigger than nuclear spacing $ \sim 10^{-10}m$ and smaller than the characteristic size of the experiment. A useful educated guess is the UCN's de Broglie wavelength
\begin{equation}
    \sigma\sim \frac{ h}{m \norm { \vec \upsilon}} \sim 10^{-8}m \,,
\end{equation}
which is approximately the same result we would get from the Heisenberg principle for the minimal value of the position uncertainty, being the velocity-dispersion $\Delta \upsilon$ for qBounce UCN about $\sim 1 \, \nicefrac{m}{s}$. In fact, from the state \eqref{initialstate} we find
\begin{equation}
    m \Delta \upsilon = \frac{\hbar}{\sqrt{2} \sigma } \to \sigma = \frac{\hbar }{\sqrt{2} \,m \Delta \upsilon} = 4.45 \times 10^{-8}\,m \,.
\end{equation}

\subsection{Newtonian Schr\"odinger problem}

At the lowest order ($c^0$), the theoretical description of the UCNs in the qBounce setting is given by the Hamiltonian $H_N$ in \eqref{HNexpr}, where spin does not play any role. Using Eq.\eqref{Hzform} and \eqref{HNexpr}, we can calculate the leading order effective Hamiltonian 
\begin{equation} \label{HNz1}
    H_N^{(Z)}= \frac{\hbar^2}{2m \sigma^2} + \frac{\vec k_\perp^2}{2m} +\frac{\hbar^2}{2m} \partial^2_Z + m a Z \, .
\end{equation}
Its spectrum $\calE^{(0)}$ is given by solving the correspondent secular equation
\begin{equation}
    \left (\frac{\hbar^2}{2m} \partial^2_Z +maZ \right ) \, \varphi (Z) =  E \, \varphi(Z) \,,
\end{equation}
with $E \equiv \calE^{(0)}-\frac{\vec k_\perp^2}{2m}- \frac{\hbar^2}{2m \sigma^2} $ . 
 The presence of the qBounce bottom mirror is simulated by setting boundary conditions at $Z=0$. Fortunately, the solution of the above equation is well-known and is given by Airy functions \cite{pitschmann2019schr}
\begin{equation}
    \varphi_n (Z) = C_n \, Ai \left (\frac{Z-Z_n}{Z_0} \right ) \,,
\end{equation}
with 
\begin{equation} \label{airyparamdef}
    C_n = \frac{Z_0^{-\frac12}}{Ai'\left(-\frac{Z_n}{Z_0}\right )} \, ,\,\,Z_0 = \left(\frac{\hbar^2}{2 m^2 a} \right)^{\frac13}\,,\,\, Z_n = \frac{E_n}{m a }\,,
\end{equation}
where the $Ai'(\zeta)$ represents the derivative of $Ai(\zeta)$ with respect to its argument $\zeta \equiv \frac{Z-Z_n}{Z_0}$.  The $E_n- $values are determined by the quantization condition obtained by setting the wavefunction at $Z=0$ to zero
\begin{equation}
    Ai \left (-\frac{E_n}{m a Z_0} \right )=0 \,.
\end{equation}
Thus, zeroth-order spectrum for UCN is
\begin{equation} \label{zerospectrum}
    \calE_n^{(0)} = \frac{\hbar^2}{2m \sigma^2}+\frac{1}{2} m \upsilon_\perp^2+ E_n \, \,\, ,\,\,\, \upsilon_\perp\equiv \frac{\norm{k_\perp}}{m} \,.
\end{equation}

\subsection{Next-to-Leading Order (NLO) Corrections}

%To derive the NLO-corrections to the spectrum \eqref{zerospectrum}, we first have to calculate the curvature tensor and therefore assume a prior spacetime, for which we choose the ERS metric structure \eqref{ERSmetric}. 
We are now ready to study the energy corrections due to NLO contributions: the first step is to integrate out the $XY-$dynamics also from the NLO Hamiltonian contribution in \eqref{HNLOfin} as
\begin{eqnarray} \label{HZNLOint}
    H^{(Z)}_{NLO} = \int d^2 X_\perp \phi^*(\vec X_\perp) \, H_{NLO} \,  \phi(\vec X_\perp) \,,
\end{eqnarray}
whose full expression is provided in \eqref{HZNLOfin}. At this point we are left with a perturbation to the Newtonian Hamiltonian \eqref{HNz1}. Since NLO terms introduce operators involving Pauli matrices, for each value of the quantum number $n$ we now have a two-dimensional eigenspace, spanned by the degenerate eigenvectors of the unperturbed problem
\begin{eqnarray} 
\langle Z| \,\varphi_n , \uparrow \rangle &=& C_n \,  Ai \left( \frac{Z-Z_n}{Z_0}\right) \left( \begin{array}{c} 1 \\ 0 \end{array}\right)\,, \\
\langle Z | \, \varphi_n, \downarrow \rangle &=& C_n \,  Ai \left( \frac{Z-Z_n}{Z_0}\right) \left( \begin{array}{c}0 \\1 \end{array}\right) . \nonumber
\end{eqnarray}
To apply standard quantum perturbation theory, we must first calculate the matrix elements $W^n_{\alpha \beta}$ of the perturbation within the degenerate unperturbed subspaces
\begin{equation} \label{WNab}
    W^{n}_{\alpha \beta} \equiv \bra{ \varphi_n ,\,\alpha } H^{(Z)}_{NLO} \ket{\varphi_n ,\,\beta} =
    \bra{ \varphi_n } (H^{(Z)}_{NLO})_{\alpha \beta} \ket{\varphi_n } \, ,
\end{equation}
where $(H^{(Z)}_{NLO})_{\alpha \beta} $ are the matrix components of the NLO Hamiltonian in the two-dimensional spin subspace with $\alpha, \beta = \, \uparrow, \downarrow$. To calculate the first-order corrections $\calE_n^{(1)}$ we therefore have to diagonalize the $W^n_{\alpha \beta}-$matrices to find their eigenvalues as solutions of the equation
\begin{equation}
     \det \left ( W^n - \calE_n^{(1)} \mathbf{1}  \right ) = 0 ,
\end{equation}
with $\mathbf{1}$ the two-by-two identity matrix. Expanding the determinant, we arrive at the secular equation
\begin{equation} \label{eigenvalueeq}
   ( W^{n}_{\uparrow\uparrow}-\calE_n^{(1)})^2 - |W^{n}_{\uparrow \downarrow}|^2=0 \, ,
\end{equation}
where we used the fact that, in our case, we have
\begin{equation} \label{HNNelemcond}
W^{n}_{\uparrow\uparrow}=W^{n}_{\downarrow\downarrow} \, , \, \, W^{n}_{\uparrow\downarrow} = (W^{n}_{\downarrow\uparrow})^* \, .
\end{equation}
Thus, the next-to-leading order corrections to the spectrum will be given by
\begin{equation}\label{corr1.1}
    \calE^{(1)}_{n,\pm}= W^{n}_{\uparrow\uparrow} \pm W^{n}_{\uparrow\downarrow} \, .
\end{equation}
To compute the $W^n_{\alpha \beta}$, we now have to calculate the averages of all the $Z-$dependent quantities appearing in Eq.\eqref{HZNLOfin}, which involve integrals of Airy functions multiplied by powers of $Z$ and $\partial_Z$. Suprisingly, this can be done analytically, by using the method reported in \cite{pitschmann2019schr}. The calculation technique involves shifting the argument of the second $Ai$ by a small quantity $\lambda$ and then taking the limit $\lambda \to 0$ after performing the integrals. We include details about this procedure in Appendix \ref{AIintegral}. \\
Assuming the following general notation for the averages over the Airy eigenstates
\begin{equation} \label{avernotation}
    \braket{O}_n = \bra{\psi _n^{(0)}} O \ket{\psi _n^{(0)}} \, ,
\end{equation}
at the end, we explicitly obtain the following mean values
\begin{equation} \label{zaverag}
\begin{cases}
\braket{\partial_Z}_n = 0 \, ,\, \, \braket{ \partial^2_Z}_n = -\frac{1}{3} \frac{Z_n}{Z_0^3} \, ,\,\, \braket{Z}_n = \frac{2}{3}Z_n \, , \\
\braket{Z^2}_n = \frac{8}{15} Z^2_n \,,\,\,
\braket{Z \,\partial_Z^2}_n = -\frac{2}{15} \frac{Z^2_n}{Z_0^3} \, , \, \, \braket{\partial^4_Z}_n = \frac{1}{5} \frac{Z_n^2}{Z_0^6}\,.
\\
\braket{Z \, \partial_Z }_n = -\frac12 \,,\,\,
\braket{Z^2 \partial_Z^2}_n = \frac37-\frac{8}{105} \frac{Z^3_n}{Z_0^3} \,.
\end{cases} 
\end{equation}
Combining formulas \eqref{corr1.1}, \eqref{zaverag}, \eqref{HZNLOfin} and neglecting constant shift terms, we obtain
\begin{eqnarray} \label{pNspectrum}
    \calE^{(1)}_{n,\pm} &=& 
    -\frac{8 m \,G M Z_n^2}{15 R^3}-
    (5-6b-5d)\frac{4 m \, G^2 M^2 Z_n^2}{15 \,c^2 R^4} \nonumber\\
    && +\frac{ m a  Z_n}{6 c^2} (\upsilon_\perp^2 + \frac{\hbar^2}{m^2 \sigma^2}+ (3-2d)\frac{ G M \sigma^2}{3 \, R^3 } )  \\
    && +\frac{ m a^2 Z_n^2}{30\,c^2} -(d+3) \frac{4 \,m G M  Z_n^2}{45 \, c^2 R^3}(\upsilon_\perp^2 + \frac{\hbar^2}{m^2\sigma^2}) \nonumber \\
    && -\frac{8 \, m a G M  Z_n^3}{105\, c^2 R^3} \pm \,\frac{\,\hbar \,\upsilon_\perp}{4 c^2} \left ((d+2)\frac{2 G M Z_n}{3 R^3}-a \right )  \nonumber .
\end{eqnarray}
Note that we grouped $v^F-$corrections \eqref{vhfermi} in the first line, while mixed $\tilde h^F_{ij}-$corrections are present in the remaining lines.\\ 

\subsection{Remarks on NLO contributions in literature}

As previously mentioned in the Introduction and in Section III for the FW transformation, NLO corrections to the Hamiltonian for fermions in weak gravitational fields are often packed in a $\nicefrac{1}{m}-$expansion. This choice can lead to inconsistent result, particularly when coupled with the additional caveat of keeping terms at most linear in $\Phi_S$: consider the case of the following first-order Hamiltonian (in natural units $c=\hbar=1$) \cite{kostelecky2021searches,ivanov2021quantum}
\begin{equation} \label{H1orderlit}
    H_{NLO}^{(lit)} = \frac{3}{4m} (\vec \sigma \times \vec p) \cdot \vec g -\frac{3}{4m} (\vec p^2 \vec g \cdot \vec z + \vec g \cdot \vec z \, \vec p^2) \,.
\end{equation}
 Already from the averages in \eqref{zaverag}, we can observe that there are pieces missing from \eqref{H1orderlit} at this level of approximation. In fact, since 
\begin{equation}
\frac{\braket{p^2_z}_n}{m^2} =
     a \braket{ z}_n \simeq \Phi_S  - \bar \Phi_S  \, ,
\end{equation}
we can identify the following terms to generate corrections of the same order of magnitude
\begin{equation} \label{orderpredict}
    {\Phi_S \frac{p^2_z}{m^2}} \sim { \Phi_S^2} \sim \frac{p^4_z}{m^4} \,.
\end{equation}
This fact directly highlights two problems
\begin{itemize}
    \item Working in the framework of a linear weak-gravity expansion, leads to the partial exclusion of $z^2-$corrections coming from $\nicefrac{1}{c^4}-$contributions to $v$ or equivalently $\Phi_S^2-$contributions in $g_{00}$. This is especially true if the additional assumption $\vec g = const$ is taken.
    \item The misleading use of the $\frac{1}{m}-$expansion parameter can also lead to erroneously neglect the first special-relativistic corrections $\propto \partial_z^4$.
\end{itemize}
Both these exclusions lead to neglecting terms whose corrections are of the same order of the one given by \eqref{H1orderlit}, and should therefore be included. \\
Those considerations strengthen our choice of a $\nicefrac{1}{c}-$expansion and the successive use of the Fermi coordinates, two prescriptions which also other authors started to adopt in recent years \cite{alibabaei2023geometric}.

\subsection{Estimation for qBounce}

Let's finally give some estimations for the effects in the qBounce context. Combining the results \eqref{zerospectrum} and \eqref{pNspectrum}, with a little algebra and the exclusion of $n-$independent terms, we get
\begin{eqnarray} \label{finalspectrum}
    \calE_{n,\pm} &=& 
    \left [1+ \frac{1}{6 c^2} (v_\perp^2 + \frac{\hbar^2}{m^2\sigma^2} + (3-2d)\frac{ \, GM \sigma^2}{3 R^3}) \right ]  E_n \nonumber\\
    && - \frac{1}{5 m} \left [ \frac{8 \, G M}{ 3 \, a^2 R^3} \nonumber -\frac{1}{6 c^2 }  - (5-6b-5d) \frac{4 \, G^2 M^2 }{ 3 a^2 c^2 R^4 }\right . \\
    && \;\;\;  \;\;\;  \;\;\; \;\; \left . + (d+3) \frac{4 \,  G M }{9 \,  c^2 a^2 R^3} (v_\perp^2 + \frac{\hbar^2}{m^2\sigma^2}) \right ]  E^2_n \\
    && + \frac{8\, G M \, E_n^3}{105 \, m a^2 c^2 R^3}  \pm \frac{\hbar \, v_\perp}{4 c^2} \left ((d+2)\frac{2 \,G M \, E_n }{3 \, m a R^3} -a \right ) .\nonumber
\end{eqnarray} 
Using the values in table \ref{tab:qBounce} and using the conservative estimate $\sigma \sim 10^{-8}\,m$, we easily see that by far the largest perturbation comes from the first term in the squared parenthesis proportional to $E_n^2$, which would be order $10^{-12}\times E_n$. Actually, that term is analogous to what we would get expanding the $\nicefrac{1}{r}-$potential around the Earth's surface to quadratic order and, in principle, it should not be considered as a part of the pN corrections but of the higher order $\norm {X} -$corrections. In our scheme, in fact, it is generated by the first term in equation \eqref{vhfermi} when considering $\nicefrac{1}{c^2}-$contributions of the Riemann tensor in Eq.\eqref{Rtransform}.\\
The consequence of these considerations are twofold:
\begin{itemize}
    \item First of all, since the terms involving the PPN parameters $b $ and $d$ are the smallest ones, it is unlikely that useful upperbounds can be put through these type of experiments.
    \item Secondly, the NLO corrections
    will not be observable in this class of experiments in the near future.
\end{itemize}

\subsection{Local$-a$ Tension}

Recently, an interesting discrepancy among the local acceleration value measured by a classical experiment $a_{cl} = 9.8049 \, \nicefrac{m}{s^2}$ and by the qBounce experiment $a_{qB} = 9.8120 \, \nicefrac{m}{s^2}$ has been reported in \cite{micko2023qbounce}. This inconsistency was observed by deducing the effective value of the acceleration by studying the transition among the energy level $n=1$ to $n=6$. The  experimentally derived value for the transition frequency $\nu_{1\to6 }$ is 
\begin{equation} \label{expnu16}
    \nu_{1\to6}^{obs} = 972.81 \,Hz\,,
\end{equation}
while the predicted value from \eqref{zerospectrum} corresponds to $972.35 \,Hz$. Since the statistical significance of this result already reached several sigmas, assuming there is no systematical flaw in the experimental derivation, one should start to search for causes of this shift. The cautious way to go here, before thinking to some new physics hint, is to take into consideration $NLO-$effects and see whether they can account for the discrepancy. \\
Therefore, we calculate $\nu_{1\to6}$ with our corrected spectrum \eqref{finalspectrum}, while fixing $b=d=1$ and letting $\sigma $ as the only “free” parameter (being the only quantity whose value could in principle lay in a range spanning several orders of magnitude). This way we find
\begin{equation}
    \nu_{1\to6}^{NLO} =( 972.35  +9.24 \, \{\sigma\}^2 
\times 10^{-22} +  \frac{7.15}{\{\sigma\}^2} \times 10^{-30}) \,Hz\,,
\end{equation}
where with $\{\sigma\}$ we are indicating just the numerical value of $\sigma$ when expressed in meters. Thus, we see that to obtain the experimental value \eqref{expnu16} we should have $\sigma = 3.94 \times 10^{-15} m$ (curiously close to the physical diameter of neutrons) or $\sigma = 2.23 \times 10^{10} m$. Such values are, however, not likely since they are way out the allowed region defined by the qBounce setting. This is just another confirmation of the fact that for NLO corrections to be relevant we have to push parameters to values which lie outside their realistic ranges for current experiments. Thus, the local-$a$ tension still awaits an explanation.

\section{Conclusions} \label{concl}

In this work, we have calculated the non-relativistic and low-curvature corrections to the Schr\"odinger equation for a ultracold neutron in a static spacetime. We have done that starting from the Dirac equation on the curved spacetime generated by the Earth's gravitional field. The whole process involves many different technicalities, like the FW transformation and the proper reference frame choice, which makes it highly non-trivial, despite the amount of literature on the subject. In fact, terms that could seem negligible at first glance end up being of the same order of the other perturbative corrections to the neutron energy spectrum, when doing things consistently. In this sense, we have seen that a post-Newtonian approach can help to avoid these difficulties.\\
Finally, we analyzed our results 
from an experimental perspective and found that, with the current level of precision, post-Newtonian corrections will not play a role in near-future observations for experiments like qBOUNCE or GRANIT, unless drastic changes to the setups, while maybe being for others. That also implies that UCN experiments may not be useful in determining deviations from GR predictions for the PPN parameters.\\
Nevertheless, the positive side is that any new tension that may appear in their measurements, like the one on the local acceleration value mentioned in the text, could be regarded as a sign of new physics, after carefully excluding any alternative origin for systematic errors.

\begin{acknowledgments}

We thank H.~Abele and M.~Pitschman for comments and discussion.
A.S. acknowledges financial support from ANID Fellowship CONICYT-PFCHA/DoctoradoNacional/2020-21201387. E.M. acknowledges financial support from Fondecyt Grant No 1230440 and ANID PIA Anillo ACT192023.\\

\end{acknowledgments}
 
\appendix

%%%%%% NEW APPENDICES %%%%%%%%%%%%

\section{Expressions for $\Gamma_\mu$}\label{gammamu}

Here we calculate the expression of the Spin-Connection $\Gamma_\mu$. Let's start by the time component 
\begin{eqnarray}
    \Gamma_t &=& \frac{1}{2} \sigma^{ab} g_{\nu\rho} \,e_a^{\;\;\nu} \, \nabla_t e_{b}^{\;\;\rho}  \\
    & =& \frac12 \sigma^{ab} \, g_{\nu\rho} \, e_a^{\;\;\nu} e_{b}^{\;\,\gamma} \{^{\;\, \rho}_{ t\,\, \gamma} \} \, , \nonumber
\end{eqnarray}
where we already used the fact that working with static metrics, nothing can depend on time. Expanding the expression for the Christoffel symbol and remembering that $g_{ti} = e_0^{\;\;i} = 0$, we have
\begin{eqnarray}
    \Gamma_t &=& \frac14 \sigma^{ab} \, g_{\nu\rho} g^{ \rho  \alpha }\, e_a^{\;\;\nu} e_{b}^{\;\,\gamma} (\partial_\gamma g_{\alpha t} - \partial_\alpha g_{\gamma t} )  \nonumber \\
    &=& \frac14 \sigma^{ab} \, e_a^{\;\;\nu} e_{b}^{\;\,\gamma} (\partial_\gamma g_{\nu t} - \partial_\nu g_{\gamma t} ) \, ,  \\
    &=& \frac12 \sigma^{ab} \, e_a^{\;\;t} e_{b}^{\;\,j} \, \partial_j g_{t t} = \frac12 \sigma^{0 I} \, e_0^{\;\;t} e_{I}^{\;\,j} \, \partial_j g_{t t} \,. \nonumber
\end{eqnarray}
Using the expressions in \eqref{staticmetric} and \eqref{timetetrad}, we finally obtain
\begin{equation} \label{GTcomplete}
    \Gamma_t = \frac{1}{2V} \sigma^{0 I} \, e_{I}^{\;\,j} \, \partial_j V^2 = \sigma^{0 I} \, e_{I}^{\;\,j} \, \partial_j V \,.
\end{equation}
The spatial components of $\Gamma_i$ can be calculated in an analogous fashion, by taking into consideration that
\begin{equation}
    \{^{\;\, t}_{ i\,\,\, j} \}=\{^{\;\, j}_{ i\,\,\, t} \}=0 \, .
\end{equation}
At the end of the day, we get 
\begin{equation} \label{GIcomplete}
    \Gamma_i = \frac{1}{2} \sigma^{KL} ( g_{mn } e_{K}^{\;\;m} \partial_ i e_{L}^{\;\;n } + e_{K}^{\;\;m} e_{L}^{\;\;n } \partial_n g_{im} ) \,.
\end{equation}
%which exploiting the (anti)simmetries can be written as
%\begin{equation} \label{GIcomplete}
 %   \Gamma_i = \frac{1}{4} \sigma^{KL} ( g_{mn } (e_{K}^{\;\;m} \partial_ i e_{L}^{\;\;n } -e_{L}^{\;\;n} \partial_ i e_{K}^{\;\;m }) + 2 e_{K}^{\;\;m} e_{L}^{\;\;n } \partial_n g_{im} ) .
%\end{equation}

\section{Foldy-Wouthuysen Transformation}\label{FWmethod}

The Foldy-Wouthuysen transformation is a technical tool to construct a connection between Dirac theories and their Schr\"odinger equivalent in the non-relativistic limit $|\vec p | << m c$. In the following, we include details on some calculations that lead to results used in the main text. Remembering that $S = -i\frac{\beta \Theta}{2  m c^2} \sim {\mathcal{O}}(c^{-1})$, we can easily find the structures of the commutators in Eq.\eqref{HFWcommut} up to the needed order of $\nicefrac{1}{c^2}$
\begin{eqnarray}
    [S, \mathcal H_D]  &=& i \Theta -i\frac{\beta}{2m c^2} [\Theta, \epsilon] -i\frac{\beta \Theta^2}{m c^2} \,, \\
    \,[S,[S, \mathcal H_D]] &=& \frac{\beta  \Theta^2}{m c^2}  - \frac{1}{4 m^2 c^4} [[\Theta, \epsilon] , \Theta] +\frac{\Theta^3}{m^2 c^4}  \,, \nonumber\\
    \,[S,[S,[S,\mathcal H_D]]] &\simeq& i\frac{\Theta^3}{2 m^2 c^4} -i \frac{\beta \Theta^4}{2m^3 c^6} \,, \nonumber \\
    \,[S,[S,[S,[S,\mathcal H_D]]]] &\simeq& \frac{\beta \Theta^4}{6 m^3 c^6} \, .\nonumber
\end{eqnarray}
Putting all of these expressions together, we obtain
\begin{eqnarray} \label{HFWafterI}
    \mathcal H^{I}_{FW} &\simeq& \beta m c^2 + \mathcal{E} +\frac{1}{2m c^2} \beta  [\Theta, \epsilon]+ \frac{1}{2m c^2} \beta \Theta^2\\
    &&+\frac{1}{8 m^2 c^4} [[\Theta,\epsilon],\Theta] -\frac{1}{3m^2 c^4}\Theta^3 - \frac{ 1}{8m^3 c^6} \Theta^4 \nonumber \,.
\end{eqnarray}
Repiting the same procedure two more times, we can completely get rid off the odd terms up to order $c^{-2}$
\begin{eqnarray} \label{HFWafterIII}
    \mathcal H^{III}_{FW} &=& \beta m c^2 +\mathcal{E} + \frac{\beta \Theta^2}{2m c^2} \\
    &&+\frac{\beta}{8 m^2 c^4} [[\Theta,\mathcal{E}],\Theta] - \frac{ \beta}{8m^3 c^6} \Theta^4 + {\mathcal{O}}(c^{-3})\nonumber \,.
\end{eqnarray}
We now just have to calculate each one of the structures appearing in the Hamiltonian above up to the relevant perturbative order. This task can be completed quite straightforwardly taking into account the expressions for even and odd operators in \eqref{epsilontheta} and the commutation rules for Dirac matrices. In fact, when calculating objects like $\Theta^2$ one should be very careful since, for example, $\sigma^{KL}$ does not simply commute with the $\alpha-$matrices
\begin{eqnarray}
    \alpha^i \sigma^{kl} \alpha^j &=& -\frac14 \gamma^i [\gamma^k,\gamma^l] \gamma^j \\
    &=& - \gamma^i \gamma^j \sigma^{kl} -\gamma^i \gamma^l \delta^{jk}- \gamma^i \gamma^k \delta^{jl} \nonumber  \\
    &=& \alpha^i \alpha^j \sigma^{kl} +\alpha^i \alpha^l \delta^{jk}+ \alpha^i \alpha^k \delta^{jl} \nonumber \,,
\end{eqnarray}
where we had to take into account that 
\begin{equation}
    \gamma^i \gamma^j = 2 \,\eta^{ij} - \gamma^j \gamma^i =
    - 2\, \delta^{ij} - \gamma^j \gamma^i \, .
\end{equation}
We remember that at this point of the calculations we are already adopting the new convention \eqref{newindnot} to simplify the reading.

\subsection{Non relativistic Hamiltonian}

After performing all the above cited calculations and adding up the pieces, we end up with
\begin{eqnarray} 
    \mathcal H^{III}_{FW} &=& \beta m c^2 + \beta m c^2 v - \frac{\hbar^4 \partial_i^4}{8 m ^3c^2} \\
    &\;\;\;-& \frac{\beta \hbar^2 }{2 m}
    \left  \{ (1+v) \partial_i ^2 + 2 \, \varepsilon_i^{\;\;j} \partial_i \partial_j + 2 \, \Gamma_i \partial_i +\partial_i \Gamma_i\nonumber \right . \\
    && \;\;\;\; +\frac12 \partial_i ( \, 2 v \, \delta_{il}+\varepsilon_l^{\;\;i}+\varepsilon_i^{\;\;l} +h_{il}) \, \partial_l   \nonumber \\
    && \;\;\;\; + \frac14 \partial_i ^2 v +\frac14 \partial_i ^2 h + i \,  \epsilon^{ijk} \Sigma^k \partial_i \Gamma_j 
     \nonumber \\
    && \;\;\;\; \left. + \frac{i}{2} \epsilon^{ijk} \Sigma^k \partial_i ( v \, \delta_{jl} +\varepsilon_{j}^{\;\;l}+ \varepsilon_{l}^{\;\;j} - h_{jl}) \, \partial_l \right \} .\nonumber
\end{eqnarray}
Exploiting the properties in \eqref{tetradcond} we obtain
\begin{eqnarray} \label{HFW3}
    \mathcal H^{III}_{FW} &=& \beta m c^2 + \beta m c^2 v - \frac{\hbar^4 \partial_i^4}{8 m ^3c^2} \\
    &\;\;\;-& \frac{\beta \hbar^2 }{2 m}
    \left  \{ (1+v) \partial_i ^2 + h_{ij }\partial_i \partial_j + 2 \, \Gamma_i \partial_i +\partial_i \Gamma_i \right .  \nonumber\\
    && \;\;\;\; + \partial_i ( \,v \, \delta_{il} + h_{il}) \partial_l + \frac{i}{2} \epsilon^{ijk} \Sigma^k \partial_i v \, \partial_j   \nonumber \\
    && \;\;\;\; + \frac14 \partial_i ^2 ( v+h) + i \,  \epsilon^{ijk} \Sigma^k \partial_i \Gamma_j \left. \right \} , \nonumber
\end{eqnarray}
which is already a quite compact form. To simplify it even more, we make use of the choice \eqref{symmtetrad}, which in our expansion scheme is always a possible one: in fact, in this case the $\Gamma_j$ reduce to \eqref{simplegammaj} and therefore
\begin{equation}
    \Gamma_j = \frac{i}{4} \epsilon^{jkl} \Sigma^j \partial_k h_{il} \, ,
\end{equation}
which when replaced into \eqref{HFW3}, defining $\tilde h_ {ij} \equiv h_ {ij} + v \delta_ {ij}$, directly lead to the final formula \eqref{finalHNLO} in the main text. 
\vspace{0.1in}

\section{Hamiltonian for ERS spacetime in Fermi coordinates} \label{HERSinFermi} 

Here, we will include the expression for the Fermi Hamiltonian \eqref{HNRsum} taking the ERS spacetime as our prior spacetime structure. The relevant quantities to use are
\begin{eqnarray} \label{vhfermiers}
    v^F &=& \frac{a Z}{c^2} + \frac{GM (\vec X^2_\perp-2Z^2)}{2\,c^2 R^3} \\
    &&+ \frac{ \, G^2 M^2}{2 \, c^4 R^4} \left [ (5d+6b-5)Z^2- (2b+3d-2) \vec X^2_\perp \right ] \nonumber \\
    h^F_{ij} &=& 
    \frac{d\, G M}{3c^2 R^3}\left( \begin{array}{ccc } 
    \; 2Y^2 -Z^2& -2X Y & XZ \\ 
    -2XY & 2X^2-Z^2 & YZ \\ 
    XZ & YZ & -\vec X^2_\perp
    \end{array}\right) 
\end{eqnarray}
which were calculated by evaluating the Riemann tensor components on the observer's worldline, setting \eqref{staticworldline}. Combining these expressions with Eq.\eqref{finalHNLO} we obtain the NLO Hamiltonian correction
\begin{widetext}
\begin{eqnarray}\label{HNLOfin}
    H_{NLO} &=& \frac{G M m }{2R^3 } (\vec X_\perp^2 -2 Z^2)   + \frac{ m \, G^2 M^2}{2 \, c^2 R^4 } \left ( (2-2b-3d) \vec X_\perp^2 - (5-6b-5d) Z^2 \right ) \\
    &&-\frac{\hbar^2  a }{2m c^2}\partial_Z  +  \frac{d-3}{6}\frac{\hbar^2 G M}{ m c^2 R^3 } (\vec X_\perp \cdot \vec \partial_\perp - 2 Z \partial_Z )  +\frac{i \hbar^2  }{4m c^2}  \left (a -(d+2) \frac{G M}{R^3} Z \right )(\sigma^X \partial_Y - \sigma^Y \partial_X ) \nonumber \\
    && + \frac{i \hbar^2 G M}{4 m c^2 R^3} \left  ((d-1)  (\sigma^X Y - \sigma^Y X) \, \partial_Z  +(2d +1 ) \, \sigma^Z (\,Y \partial_X - X \partial_Y ) \right )   \nonumber \\
    &&  - \frac{\hbar^2 a  }{2 m c^2} Z \partial_i ^2+(2d-3) \frac{\hbar^2 G M \vec }{12 m c^2 R^3} X^2_\perp \partial^2_Z - (4d+3) \frac{\hbar^2 G M}{ 12 m c^2 R^3} ( X^2 \partial^2_Y + Y^2 \partial_X ^2 ) - \frac{\hbar^2 G M }{4 m c^2 R^3 } (X^2 \partial_X^2 + Y ^2 \partial_Y^2 ) \nonumber \\
    &&  + \frac{\hbar^2 G M }{2 m c^2 R^3} Z^2 \partial_Z ^2 + (d+3) \frac{\hbar^2 G M }{6 m c^2 R^3} Z^2 \vec \partial_\perp^2 -\frac{d}{3}\frac{ \hbar^2 G M}{ m c^2 R^3} ( X Z \, \partial_X \partial_Z + Y Z \, \partial_Y \partial_Z -2 X Y\, \partial_X\partial_Y ) - \frac{\hbar^4 \partial_i^4}{ 8 m^3 c^2}\,. \nonumber 
\end{eqnarray}
\end{widetext}
which is sorted in order of increasing derivatives. Note that the first term in the above expression is just the second order contribution to the expansion of the classical Newtonian $\nicefrac{1}{r}\,-$potential.\\
At this point, integrating out the $XY-$dynamics as in \eqref{HZNLOint}, we get the final perturbation form to calculate the spectrum's corrections 
\begin{eqnarray}\label{HZNLOfin}
    H^{(Z)}_{NLO} &=& \left ((2d-3) \,  \frac{G M \sigma^2}{6 R^3} +\frac{ \vec k_\perp^2}{2m^2 } + \frac{ \hbar^2}{2 m^2 \sigma^2}  \right )\frac{\hbar^2 \partial_Z^2}{2m c^2} \nonumber \\
    &\;\;\;+&  (\vec k^2 _\perp + \frac{\hbar^2}{\sigma^2 }  ) \frac{a Z}{2 m c^2 } - \left ( 1+ \frac{(d+3) \hbar^2}{6 m^2 c^2 \sigma^2}(\vec k^2 _\perp + \frac{\hbar^2}{\sigma^2 }  ) \right . \nonumber \\
    &\;\;\;-& \left . (5-6b-5d) \frac{G M}{2 c^2 R } \right )\frac{G M m}{R^3}Z^2 - \frac{\hbar^4 \partial_Z^4}{ 8 m^3 c^2}   \\
    &\;\;\;-& \frac{ \hbar^2 a }{ 2 m c^2} (1+Z\partial_Z) \partial_Z +\frac{ \hbar^2  G M }{m c^2 R^3} (Z \partial_Z +\frac{Z^2 \partial_Z^2}{2} )  \nonumber \\
    &\;\;\;+& \frac{\hbar }{4mc^2}\left ( (d+2) \frac{GM Z}{R^3} -a \right) (\sigma^X k^Y  - \sigma^Y k^X  ) \, . \nonumber 
\end{eqnarray}
from which, being interested in transition energies, we already removed the $Z-$independent terms, since their effect would get cancelled in the differences between energy levels.\\
To not include here even more large formulas, we avoid to write the expressions for the single matrix components $W^n_{\alpha \beta}$ defined in \eqref{WNab}. Their calculation is, in fact, trivial starting from \eqref{HZNLOfin} and exploiting the relations \eqref{zaverag}.

%\section{Spectrum Corrections Details}
%\begin{eqnarray} 
%    \calE^{(1)}_{n,\pm} &=& 
%    -\frac{8 m \,G M z_n^2}{15 R^3}+
%    (1+2b+3d)\frac{4 m \, G^2 M^2 z_n^2}{5 \,c^2 R^4}+ \nonumber\\
%    && +\frac{z_n}{3mc^2} (a - \frac{\hbar^2}{4 m^2 z_0^3})(k_\perp^2 + \frac{\hbar^2}{\sigma^2}) - \frac{d \, G M \hbar^2 \sigma^2 z_n}{18 \, m c^2 R^3 z_0^3} \nonumber \\
%    && -(d+3) \frac{4 \,G M z_n^2}{45 \,m c^2 R^3}(k_\perp^2 + \frac{\hbar^2}{\sigma^2}) \nonumber\\
%    &&+
%    \frac{ \hbar^2 z_n^2}{15\, m c^2 z_0^3} (a - \frac{3 \hbar^2}{8 m^2 z_0^3}) \nonumber \\
%    && -\frac{4 d \,G M \hbar^2 z_n^3}{105\, m c^2 R^3 z_0^3}- \frac{2 d \, \hbar^2 G M }{7 \,m c^2 R^3} \nonumber \\
%    && \pm \frac{\hbar}{4mc^2} \left (\frac{2d+4}{3}\frac{G M z_n}{R^3}-a \right ) \norm{k_\perp}
%\end{eqnarray}

\section{Integrals of Airy functions} \label{AIintegral}

The problem we want to discuss in this appendix is the one related with calculating the analytic form of integrals $I[O] \equiv \braket{O}_n$ of the type
\begin{equation} \label{IOdef}
    I[O(Z)] =  \int_{0}^{\infty} dZ \,  Ai \left (\frac{Z-Z_n}{Z_0} \right ) O(Z) \,  Ai \left (\frac{Z-Z_n}{Z_0} \right )  ,
\end{equation}
where $O(Z)$ represents here a generic operator depending on $Z$ and acting on the second Airy function. Note that we will always consider $Z=0$ as a starting point for the integration since we are assuming no “floor-leakage", which would be equivalent to the addition of a Heaviside function $\theta(Z)$ in the wavefunctions. \\
We follow the strategy outlined in \cite{pitschmann2019schr}. First of all we make the change of variable $\zeta \equiv \frac{Z-Z_n}{Z_0}$ for integral \eqref{IOdef}
\begin{equation}
    I[O] = Z_0 \int_{-\frac{Z_n}{Z_0}}^{\infty} d\zeta Ai(\zeta) O(Z_n + Z_0 \,\zeta ) Ai(\zeta) \, ,
\end{equation}
At this point, the strategy is to introduce an infinitesimal shift $\lambda$ in the argument of the second Airy function, that depending on the form of $O$ will allow to easily realize the integral, so that at the end we can take the $\lambda\to0$ again
\begin{equation}
    I_\lambda[Z \,\partial^2_Z] = Z_0 \int_{-\frac{Z_n}{Z_0}}^{\infty} d\zeta Ai(\zeta) O(Z_n + Z_0 \,\zeta ) Ai(\zeta-\lambda) \, .
\end{equation}
As an example, let's consider the case of $O= Z \,\partial^2_Z$. Observing that $\partial_Z = \frac{1}{Z_0} \partial_\zeta$ we have
\begin{equation}
    I_\lambda[Z \,\partial^2_Z] = \frac{1}{Z_0} \int_{-\frac{Z_n}{Z_0}}^{\infty} d\zeta Ai(\zeta) (Z_n + Z_0 \,\zeta ) \partial_\zeta^2 Ai(\zeta-\lambda) \, .
\end{equation}
Being $\partial_{\zeta-\lambda}= -\partial_\lambda= \partial_\zeta$, we have
\begin{equation}
    I_\lambda[Z \,\partial^2_Z] = \frac{1}{Z_0} \partial_\lambda^2 \int_{-\frac{Z_n}{Z_0}}^{\infty} d\zeta Ai(\zeta) (Z_n + Z_0 \,\zeta ) Ai(\zeta-\lambda) \, ,
\end{equation}
which reduces to
\begin{eqnarray}\label{Ilambda1}
    I_\lambda[Z \,\partial^2_Z] &=& \frac{Z_n}{Z_0} \partial_\lambda^2 \int_{-\frac{Z_n}{Z_0}}^{\infty} d\zeta Ai(\zeta) Ai(\zeta-\lambda) \\
    && +\partial_\lambda^2 \int_{-\frac{Z_n}{Z_0}}^{\infty} d\zeta Ai(\zeta) \, \zeta Ai(\zeta-\lambda) \nonumber \\
    &=&  \partial_\lambda^2 \left ( \{ \frac{Z_n}{Z_0}\}_\lambda +  \{ \zeta \} _\lambda \right )  \, , \nonumber
\end{eqnarray}
where we introduced the general notation
\begin{equation}
    \{ P (\zeta) \}_\lambda= \int_{-\frac{Z_n}{Z_0}}^{\infty} d\zeta Ai(\zeta) \, P (\zeta ) Ai(\zeta-\lambda)\,.
\end{equation}
For these shifted $\zeta-$integrals we can use formulas (A31) and (A37) from \cite{pitschmann2019schr}, which remembering that $Ai(-\frac{Z_n}{Z_0})=Ai(\infty)= Ai'(\infty) =0$, lead to
\begin{eqnarray}
    \{ 1 \}_\lambda &=& \frac{1}{\lambda} Ai' \left (-\frac{Z_n}{Z_0} \right ) Ai \left (-\frac{Z_n}{Z_0}- \lambda \right ) , \\
    \{ \zeta \}_\lambda &=& -\frac{2+ \lambda^2 (-\frac{Z_n}{Z_0})}{\lambda^3} Ai'\left (-\frac{Z_n}{Z_0} \right ) Ai \left (-\frac{Z_n}{Z_0}-\lambda \right ) \nonumber \\
    &&-\frac{2}{\lambda^2} Ai '\left (-\frac{Z_n}{Z_0} \right ) Ai ' \left (-\frac{Z_n}{Z_0}-\lambda \right )\,. \nonumber
\end{eqnarray}
After replacing the above expressions back into \eqref{Ilambda1}, the following steps are to take the $\lambda -$derivatives and carefully expand all the shifted $Ai-$functions for small values of $\lambda$, as shown in $(A8)$ of \cite{pitschmann2019schr}, where the authors also take in consideration that Airy's functions satisfy
\begin{equation}
    \partial^2_{\zeta} Ai (\zeta) = \zeta \, Ai (\zeta) \, .
\end{equation}
At the end of the day, if one does things correctly, it should end up with an expression for which it is easy to take the $\lambda \to 0 $ limit, obtaining
\begin{eqnarray}
    \partial_\lambda^2\{ 1 \}_\lambda|_{\lambda\to0} &=&  \frac{1}{3} (-\frac{Z_n}{Z_0}) (Ai' (-\frac{Z_n}{Z_0})) ^{2} , \\
    \partial_\lambda^2 \{ \zeta \}_\lambda |_{\lambda\to0} &=& \frac{1}{5} (-\frac{Z_n}{Z_0})^2 (Ai' (-\frac{Z_n}{Z_0})) ^{2}  \,. \nonumber
\end{eqnarray}
Putting all together, we finally find that
\begin{eqnarray}
    I_{\lambda\to0}(Z \partial^2_Z) &=& \frac15 (\frac{Z_n}{Z_0})^2 (Ai' (-\frac{Z_n}{Z_0})) ^{2} \nonumber \\
    &&- \frac13 (\frac{Z_n}{ Z_0})^2 (Ai' (-\frac{Z_n}{Z_0})) ^{2} \nonumber\\
    &=& -\frac{2}{15}(\frac{Z_n}{ Z_0})^2 (Ai' (-\frac{Z_n}{Z_0})) ^{2} \,,
\end{eqnarray}
from which we can derive directly
\begin{equation}
    \braket{Z \partial^2_Z }_n = C^2_n \, I_{\lambda\to0}[Z \partial^2_Z] = -\frac{2}{15} \frac{Z_n^2}{ Z_0^3} \, .
\end{equation}
All the other mean values $\braket{\, } _n$ in the main text can be obtained with an analogous procedure.

% The \nocite command causes all entries in a bibliography to be printed out
% whether or not they are actually referenced in the text. This is appropriate
% for the sample file to show the different styles of references, but authors
% most likely will not want to use it.

\bibliography{apssamp}% Produces the bibliography via BibTeX.

\end{document}